\documentclass[10pt,conference]{IEEEtran}
\IEEEoverridecommandlockouts

\usepackage{cite}
\usepackage{amsmath,amssymb,amsfonts}
\usepackage{algorithmic}
\usepackage{graphicx}
\usepackage{textcomp}
\usepackage{xcolor}
\usepackage{hyperref}
\usepackage{array}
\usepackage{booktabs}
\usepackage{url}
\usepackage{listings}
\usepackage{xcolor}

\usepackage{caption}
\lstdefinestyle{pythonstyle}{
    language=Python,
    basicstyle=\ttfamily\scriptsize,
    keywordstyle=\color{blue!65!black},
    stringstyle=\color{teal!65!black},
    numbers=none,
    backgroundcolor=\color{gray!3},
    frame=single,
    rulecolor=\color{gray!25},
    framesep=3pt,
    breaklines=true,
    showstringspaces=false,
    columns=fullflexible,
    captionpos=t,
    belowcaptionskip=15pt,
    aboveskip=15pt,
    belowskip=5pt
}

\lstdefinestyle{bashstyle}{
  language=bash,
  basicstyle=\ttfamily\footnotesize,
  keywordstyle=,
  breaklines=true,
  columns=fullflexible,
  keepspaces=true,
  aboveskip=3pt,
  belowskip=3pt,
  numbers=left,
  numberstyle=\tiny\color{gray},
  numbersep=5pt,
  xleftmargin=1.5em,        % room for the numbers inside the column
  xrightmargin=0pt,
}

\def\BibTeX{{\rm B\kern-.05em{\sc i\kern-.025em b}\kern-.08em
    T\kern-.1667em\lower.7ex\hbox{E}\kern-.125emX}}

\begin{document}

\title{SIEVE: A Parameterized GitHub Corpus Builder\\
for Software Engineering Research}

\author{
    \IEEEauthorblockN{
        \begin{tabular}[t]{c@{\hspace{1.5cm}}c@{\hspace{1.5cm}}c}
        Musfiqur Rahman & Mahi Begoug & Emad Shihab \\
        \small musfiqur.rahman@mail.concordia.ca & \small mahi.begoug@concordia.ca & \small emad.shihab@concordia.ca
        \end{tabular}
    }
    \IEEEauthorblockA{
        \textit{Department of Computer Science and Software Engineering}\\
        \textit{Concordia University}\\
        Montreal, Canada
    }
}

\maketitle

% ── Abstract ──────────────────────────────────────────────────────────────────
\begin{abstract}
Software engineering research increasingly relies on large, richly annotated corpora of real-world code. Still, researchers typically assemble these with ad-hoc scripts or static, pre-built datasets that offer no control over corpus composition. This leaves a gap for a reusable tool that goes below the repository level and builds annotated, function- and class-level code corpora under researcher-specified criteria. We present \textsc{SIEVE} (Software Ingestion \& Extraction for Verifiable Evaluation), an open-source, parameterized GitHub corpus builder that, unlike repository-level mining tools, works at code granularity: it locates individual functions, methods, and classes in repositories and extracts them as records, giving direct control over corpus construction based on several criteria, including widely used programming languages (Python, Java, JavaScript, C++), repository quality, and creation date.
To support software engineering research and LLM evaluation on real-world code, \textsc{SIEVE} extracts CodeSearchNet- and OpenClassGen-compatible records, each annotated with 22 structural metrics and AST-derived features, and optionally with a CodeBERT-based P(AI-generated) score (F1\,=\,0.9478, AUROC\,=\,0.9902), alongside the third-party dependencies declared in each repository's package manifest. 
These annotations enable researchers to conduct structural analyses, investigate code dependencies, assess AI-generated code, and construct customized benchmarks for evaluating LLMs, including under temporal constraints. A manual validation of 400 records in four languages and two granularities shows 100\% accuracy in code metrics and import detection, 99.0\% in source completeness, and 99.5\% in metadata accuracy. \textsc{SIEVE} is publicly available with a live demo at \url{https://mrahman2025-sieve.hf.space}.
\end{abstract}


% ── I. Introduction ───────────────────────────────────────────────────────────
\section{Introduction}
\label{sec:intro}

Mining Software Repositories (MSR) research increasingly depends on large, richly annotated corpora of real-world source code drawn from open-source projects on GitHub~\cite{kalliamvakou2014promises,cosentino2017systematic}. Such corpora have long supported a wide range of research questions, such as mining GitHub data to build statistical language models of source code~\cite{allamanis2013mining}, studying the naturalness of software~\cite{hindle2016naturalness,rahman2019natural}, and studying how programming language choice relates to code quality~\cite{ray2014large}. More recently, they have enabled the evaluation of large language models (LLMs) on code generation and detection~\cite{chen2021evaluating,austin2021program,rahman2025beyond,rahman2024automatic}, the comparison of LLM-generated and human-written code~\cite{wang2025beyond,cotroneo2025human}, and analyses of dependency and library usage across ecosystems~\cite{kikas2017structure,decan2019empirical,latendresse2025robust}. Each use case places different demands on the corpus: some require multiple programming languages~\cite{ray2014large,decan2019empirical}, others a specific granularity of source code, such as functions~\cite{chen2021evaluating,husain2019codesearchnet}, classes~\cite{du2024evaluating,rahman2025beyond}, or files~\cite{kocetkov2022stack}, and some precise control over when a repository was created relative to a reference date~\cite{riddell2024quantifying,roberts2024cutoff}.

In practice, collecting such corpora from GitHub is laborious: API rate limits and result caps make large-scale sampling slow and incomplete~\cite{dabic2021sampling,gousios2012ghtorrent}, and few public tools exist for extracting information from repositories~\cite{spadini2018pydriller}, so researchers often resort to study-specific scripts 
When they reuse existing infrastructure instead, it typically covers only part of the need (Section~\ref{sec:related}).
Specialized GitHub-mining tools, including SEART~GHS~\cite{dabic2021sampling} and GHTorrent~\cite{gousios2013ghtorent}, support the sampling and filtering of repositories; however, they do not perform source-code extraction. 
Furthermore, existing static code datasets, such as CodeSearchNet~\cite{husain2019codesearchnet}, OpenClassGen~\cite{rahman2025openclassgen}, and The Stack~\cite{kocetkov2022stack}, provide pre-collected code corpora but do not allow researchers to explicitly control repository creation dates during dataset construction. 
Hence, the lack of a unified framework increases research costs and the risk of errors, as researchers repeatedly re-implement GitHub repository discovery, repository quality filtering, multilingual parsing, and metric computation from scratch for each study, and reproducing repository-based studies can be difficult~\cite{gonzalez2012reproducibility}. 
In addition, this constraint is particularly restrictive for tasks that demand fine-grained temporal control. 
For instance, assessing LLMs on code from repositories created after their training cutoff necessitates identifying and curating such repositories. 
Riddell et al.~\cite{riddell2024quantifying} report that up to 20.8\% of MBPP and 18.9\% of HumanEval solutions appear in open pre-training corpora such as The Stack, and that models perform significantly better on problems whose solutions appear in the training data. These findings highlight the need for reproducible corpus construction with temporal filtering, which can help reduce contamination risks in LLM evaluation. 

In this paper, we present \textsc{SIEVE} (\textbf{S}oftware \textbf{I}ngestion \& \textbf{E}xtraction for \textbf{V}erifiable \textbf{E}valuation), an open-source, parameterized GitHub corpus builder. Unlike repository-level mining tools, it operates at code granularity: it locates functions, methods, and classes in the source files of the selected repositories and returns each as a record with its file path, line span, and commit SHA, from which a commit-pinned GitHub permalink can be built. \textsc{SIEVE} gives researchers direct control over corpus composition along the dimensions that matter most for SE research, including programming language, code granularity, repository quality, and repository creation date.

From an architectural perspective, our tool integrates several key components into a unified pipeline, including repository discovery, multilingual code extraction, deduplication, structural metric computation, AI-generated-code scoring, dependency extraction, and export to CodeSearchNet- and OpenClassGen-compatible formats~\cite{husain2019codesearchnet,rahman2025openclassgen}. 

We evaluate \textsc{SIEVE} along three dimensions: (i) \textit{extraction correctness}, through a manual audit of 400 extracted records spanning four programming languages and two code granularities, assessing source completeness, metadata accuracy, import detection, and structural metric correctness, with accuracy between 99.0\% and 100\% across these dimensions; (ii) \textit{predictive performance}, by evaluating the integrated CodeBERT-based AI-generated-code classifier on a held-out test set of 11,419 samples, which yields an F1-score of 0.9478 and an AUROC of 0.9902; and (iii) \textit{perceived usability}, through an anonymous survey of 26 SE researchers (Section~\ref{sec:evaluation}).

\noindent
\textbf{Open source tool and documentation.}
All resources related to \textsc{SIEVE} are publicly available. See the~\nameref{sec:data-availability} section.

% % ── II. Background & Related Works ──────────────────────────────────────────
\section{Background \& Related Works}
\label{sec:related}

This section reviews existing support for building code corpora: tools for discovering and mining GitHub repositories, pre-built code datasets and benchmarks, and detectors of AI-generated code. For each, we note its limitations and how \textsc{SIEVE} complements it.

\textbf{GitHub repository discovery and mining tools.} Several tools have been developed to support different stages of GitHub repository mining. SEART GHS~\cite{dabic2021sampling} lets researchers sample repositories by metadata such as stars, commits, and creation or last-commit date ranges~\cite{seartghsweb}. While SEART GHS focuses on repository discovery and selection, GHTorrent~\cite{gousios2013ghtorent} and GH Archive~\cite{gharchiveArchive} provide historical GitHub event data. At a finer granularity, PyDriller~\cite{spadini2018pydriller} mines repositories at the commit level to support software evolution studies. Complementing these discovery and mining capabilities, Reaper~\cite{munaiah2017curating} introduced an approach for identifying engineered software projects among GitHub repositories. These tools stop at repository metadata, event streams, or commit history: none extracts function- or class-level code records, computes code metrics, or annotates code with an AI-generated-code score. \textsc{SIEVE} builds on this line of work by offering an engineered-project filter for repository selection and adding the extraction and annotation stages that follow discovery.

\textbf{Code corpora and benchmarks.} Existing code datasets serve different research purposes. CodeSearchNet~\cite{husain2019codesearchnet} provides approximately six million functions across six programming languages for code search, while The Stack~\cite{kocetkov2022stack} offers a large corpus of permissively licensed source code for LLM pre-training. Moving beyond function-level code, ClassEval~\cite{du2024evaluating} introduces 100 manually constructed Python classes for evaluating class-level code generation, which motivates the class-level records that \textsc{SIEVE} produces in the OpenClassGen-compatible schema~\cite{rahman2025openclassgen}. Despite their contributions, these datasets and benchmarks are released as predefined collections rather than configurable corpus construction tools, which limits researchers' ability to customize repository selection criteria, code granularity, and temporal constraints during data collection. In contrast, \textsc{SIEVE} provides a parameterized pipeline for constructing function- and class-level corpora from GitHub repositories across four programming languages, with configurable quality and temporal filters.

\textbf{Detecting AI-generated code.} Nguyen et al.~\cite{nguyen2023snippet} introduce GPTSniffer, a CodeBERT-based classifier that distinguishes human-written from ChatGPT-generated Java code, reporting up to 0.96 accuracy on paired snippets from a programming textbook and its ChatGPT-generated counterparts. GPTSniffer is a standalone, single-language artifact, so researchers must run it as a separate step on code they have already collected. \textsc{SIEVE} ships a CodeBERT-based classifier in the same spirit, but it covers four languages, is trained against nine contemporary LLMs, and scores every extracted record automatically alongside its other annotations.

\textbf{Positioning \textsc{SIEVE}.} Table~\ref{tab:comparison} compares \textsc{SIEVE} with the closest of these tools; static datasets are excluded because researchers consume them as released rather than run them. The tools offer complementary capabilities, but none integrates repository discovery, multilingual code extraction, structural analysis, dependency extraction, and AI-generated-code detection into one reproducible, customizable corpus construction pipeline, which is the gap \textsc{SIEVE} addresses.

\begin{table}[t]
\caption{Capability comparison of \textsc{SIEVE} with existing tools. Extraction means function- or class-level records.}
\label{tab:comparison}
\centering
\small
\setlength{\tabcolsep}{3pt}
\begin{tabular}{@{}lcccccc@{}}
\toprule
\textbf{Tool}
 & \rotatebox{90}{\textbf{Discovery \& filtering}}
 & \rotatebox{90}{\textbf{Creation-date filter}}
 & \rotatebox{90}{\textbf{Multi-lang. extraction}}
 & \rotatebox{90}{\textbf{Struct. metrics}}
 & \rotatebox{90}{\textbf{LLM-code detection}}
 & \rotatebox{90}{\textbf{Dependency extraction}} \\
\midrule
SEART GHS~\cite{dabic2021sampling}    & \checkmark & \checkmark &            &            &            &            \\
GHTorrent~\cite{gousios2013ghtorent}  & \checkmark &            &            &            &            &            \\
PyDriller~\cite{spadini2018pydriller} &            &            &            &            &            &            \\
GPTSniffer~\cite{nguyen2023snippet}   &            &            &            &            & \checkmark &            \\
\midrule
\textsc{SIEVE} (this work)            & \checkmark & \checkmark & \checkmark & \checkmark & \checkmark & \checkmark \\
\bottomrule
\end{tabular}
\end{table}

% ── III. Tool Architecture ───────────────────────────────────────────────────
\section{Tool Architecture}
\label{sec:design}

\textsc{SIEVE} is an open-source tool, implemented in Python~3.11+ and released under the MIT license~\cite{githubGitHubMrsumitbdsieve}. It exposes three interchangeable interfaces, (i) a Typer command-line interface, (ii) a Python API, and (iii) a Streamlit web application, which all call the same pipeline orchestrator.

\begin{figure}[t]
\centerline{\includegraphics[width=\columnwidth]{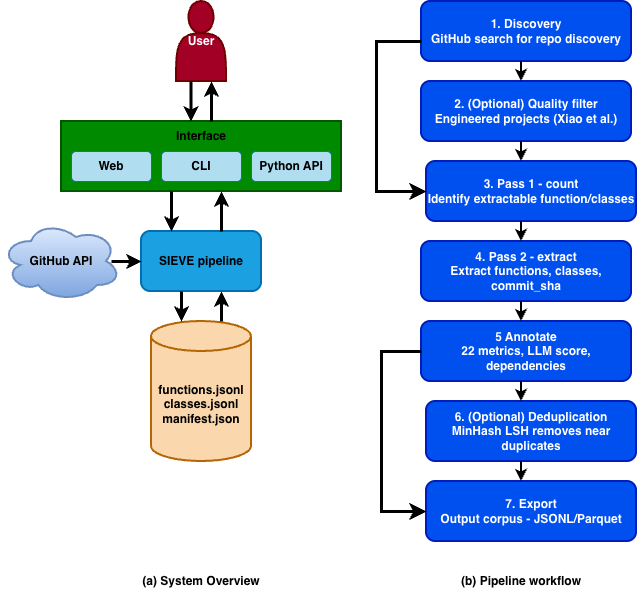}}
\caption{Overview of the \textsc{SIEVE} pipeline.}
\label{fig:architecture}
\end{figure}

Fig.~\ref{fig:architecture} gives an overview of the architecture. The orchestrator composes four modules, each described in the following subsections: (A) \emph{Repository Discovery}, which finds and filters candidate GitHub repositories; (B) \emph{Code Extraction}, which parses cloned repositories into function- and class-level records; (C) \emph{Corpus Annotation}, which adds code metrics, AST features, dependency information, and an AI-generated-code score; and (D) \emph{Deduplication and Export}, which removes near-duplicates and serializes the corpus. Section~\ref{sec:usage} provides the usage guide.

Three design choices keep the modules loosely coupled and the tool easy to extend. First, a single validated configuration object (a Pydantic model, \texttt{SIEVEConfig}) is the only input of the orchestrator \texttt{run\_pipeline}, which acts as a facade over the modules, so the three interfaces contain no pipeline logic. Second, modules exchange only plain record data classes and can be tested in isolation. Third, expensive stages (engineered-project filtering, deduplication, AI-generated-code scoring, full-AST export) are opt-in; failures are isolated, so that a repository or file that cannot be processed is skipped and reported instead of aborting the run; and progress is reported through a callback that decouples the pipeline from the interface.

\subsection{Configurable Repository Discovery}
\label{sec:discovery}

This module selects the candidate repositories of a corpus. Its inputs are the language, the creation-date range, minimum star and contributor counts, an optional repository cap, and an optional engineered-project switch. It queries the GitHub Search API~\cite{githubDocsSearchRepos} through PyGitHub with the \texttt{language:}, \texttt{created:}, and \texttt{stars:} qualifiers, excludes forks and archived repositories, and applies the contributor threshold locally using each repository's contributor count. We select by the \texttt{created:} qualifier rather than the more commonly used \texttt{pushed:} qualifier~\cite{kalliamvakou2014promises}, since the latter reflects recent activity and not when the code of a repository was written. Creation date bounds the repository itself but not the provenance of its content, because a new repository may contain code copied from older sources.
When enabled, engineered-project filtering applies the license, release, contributor, and pull-request and issue criteria of Xiao et al.~\cite{xiao2026self}, in the spirit of Reaper~\cite{munaiah2017curating}: it excludes repositories without a suitable software license, without releases, or with fewer than two contributors and, among the remaining candidates, those in the bottom quartile for pull requests and issues. The output is a metadata snapshot per repository (name, URL, stars, contributors, last commit date, default branch, license, topics, and collection time). Discovery is implemented as a generator that yields repositories lazily and stops at the requested cap, which avoids spending rate-limited API calls on repositories that are not needed.

\subsection{Code Extraction}
\label{sec:extraction}

This module turns each repository into function- and class-level records. Its inputs are the repository, the target language, the requested granularities (function/method level and class level), and optional per-repository record caps. Each repository is shallow-cloned (\texttt{--depth 1}) into a temporary directory that is removed after processing. Source files are selected by language-specific extensions, and test files and, for JavaScript, minified or vendored files are skipped. Each file is parsed with the tree-sitter grammar~\cite{treesitter,githubGitHubTreesitterpytreesitter} of the target language (Python, Java, JavaScript, or C++), and a recursive traversal of the syntax tree locates function, method, and class nodes by node type, not by text patterns.

\textbf{Function/method records} follow the CodeSearchNet schema~\cite{husain2019codesearchnet} and contain the function name, parameters, return annotation, docstring, full source, signature (header without body), used imports, decorators, line span, and, for methods, the parent class.

\textbf{Class records} follow the ClassEval~\cite{du2024evaluating} and OpenClassGen~\cite{rahman2025openclassgen} schema and contain the class name, parent classes, docstring, full source, skeleton (method signatures with empty bodies), used imports, method names and count, constructor flag, decorators, and line span.
Both record types also store the repository, file path, language, and the commit SHA at extraction time, which gives version-stable GitHub permalinks.

Extraction runs in two passes. When a corpus size cap is set, a lightweight first pass counts the extractable functions and classes of each repository, and these counts determine a proportional per-repository allocation, so that a few large repositories do not dominate the corpus. The second pass extracts records within these allocations and buffers a limited surplus, which refills any shortfall caused by repositories that yield fewer records than allocated. A separate extractor function exists for each language behind one dispatch function, and all extractors emit the same two record types, so adding a language requires registering its file extensions and grammar and writing its extractor (plus its metric node-type tables) without changing the other modules. A file that cannot be read or parsed is skipped and logged.

\subsection{Corpus Annotation}
\label{sec:annotation}
This module is responsible for automatically annotating the records collected by the extraction module. It adds code structure metrics, AST features, and an optional AI-generated-code score to each record, and the declared dependencies to each repository; these four categories are described in turn below.
Each annotator is a separate component that reads only the source code of a record (or the repository, for dependencies), so annotators can be enabled, changed, or replaced independently.

\textbf{Code structure metrics.} \textsc{SIEVE} computes 22 cross-language metrics from the tree-sitter AST of each snippet, spanning four dimensions: seven size and comment metrics (lines of code, source lines, logical lines, and comment, multi-line comment, and blank line counts, plus the comment ratio); two control-flow metrics (McCabe's cyclomatic complexity~\cite{mccabe1976complexity} and maximum nesting depth); twelve Halstead metrics~\cite{halstead1977elements} (distinct and total operators and operands, vocabulary, length, calculated length, volume, difficulty, effort, time, and estimated bugs); and the Maintainability Index~\cite{oman1992metrics} in the formulation used by Radon~\cite{radondocs}. Complete definitions are available in the tool documentation~\cite{sieveMetricDocs}.

\textbf{Dependency information.} Declared direct dependencies are parsed once per repository from ecosystem-specific manifests, using a parser suited to each format: JSON parsing for \texttt{package.json} and \texttt{vcpkg.json}, TOML parsing for \texttt{pyproject.toml}, and pattern matching for \texttt{requirements.txt}, \texttt{pom.xml}, Conan files, and CMake \texttt{find\_package} calls. Each dependency stores its name, version constraint, and kind (main, development, or optional), and the list is saved with the repository metadata in the run manifest, linked to records through the repository name.

\textbf{AST representation.} Each record stores three summary fields computed from the tree-sitter parse of its own source (tree depth, node count, and a node-type histogram) and, optionally, the full tree as nested JSON objects holding the node type, the text of leaf nodes (truncated to 80 characters), and the children. Every language is parsed through the same tree-sitter interface, so the representation has the same shape across languages, although node-type names are grammar-specific. The summary fields are cheap and small enough to be stored by default, whereas the full tree dominates file size and is therefore opt-in.

\textbf{AI-generated-code score.} Each record can optionally be scored with $P(\text{AI-generated})$ by a CodeBERT-based classifier~\cite{feng2020codebert}, whose weights are downloaded lazily from the HuggingFace Hub on first use and applied in batches. The pipeline depends only on the classifier's \texttt{load} and \texttt{score\_batch} methods and treats a scoring failure as non-fatal, so an alternative detector can be substituted without changing other modules, and a model failure does not abort a corpus build.

\subsection{Deduplication and Export}
\label{sec:export}
This module post-processes and serializes the corpus. Optional near-duplicate removal uses MinHash~\cite{broder1997resemblance} with locality-sensitive hashing, as implemented in the datasketch library~\cite{datasketch}. Each snippet is tokenized with string literals normalized, hashed into a 128-permutation MinHash signature, and dropped if the index already holds an entry whose estimated Jaccard similarity exceeds the configurable threshold (default 0.8); the first occurrence of each cluster is kept. Records are then exported as JSONL, Parquet, or both, with one file per granularity, together with a \texttt{manifest.json} that stores the full configuration, per-repository metadata, and summary counts, which allows a run to be reproduced. Export is a single function over the record lists, so adding another output format only requires adding one writer.

% ── IV. Tool Usage Scenario ──────────────────────────────────────────────────
\section{Tool Usage Scenario}
\label{sec:usage}

\textsc{SIEVE} exposes the same underlying pipeline (Section~\ref{sec:design}) through three interchangeable interfaces: (i) a command-line tool, (ii) a Python API, and (iii) a web application.

\textsc{SIEVE} is open source under the MIT license (\url{https://github.com/mrsumitbd/sieve}) and is installed from source:
\begin{lstlisting}[style=bashstyle]
git clone https://github.com/mrsumitbd/sieve.git && cd sieve
pip install -r requirements.txt && pip install -e .
\end{lstlisting}
\noindent
The editable install registers the \texttt{sieve} console command used in this section.

We illustrate all three usages with a single scenario: \textit{a researcher building a corpus of engineered-quality Python repositories created between June~1, 2024 and June~1, 2025, with at least 50 stars and 5 contributors, capped at 20 repositories, 30 functions, and 30 classes, and annotated with LLM-generation scores.}

\textbf{Command line.} The CLI is intended for headless, scripted, and cluster runs. It is invoked as \texttt{sieve run [options]}, either with individual flags or with a JSON configuration file (\texttt{--config}) that makes a run reproducible from a single artifact and takes precedence over any flags. The main options are:

\begin{small}
\renewcommand{\arraystretch}{1.1}
\noindent
\begin{tabular}{@{}>{\raggedright\arraybackslash}p{0.37\columnwidth}@{\hspace{6pt}}>{\raggedright\arraybackslash}p{0.57\columnwidth}@{}}
\hline
\textbf{Option} & \textbf{Description}\\
\hline
\texttt{--language} & Python, Java, JavaScript, or C++ (required).\\
\texttt{--start-date} & Earliest repository creation date, \texttt{YYYY-MM-DD} (required).\\
\texttt{--end-date} & Latest creation date (default: one month before today).\\
\texttt{--min-stars} & Minimum stars (default: 10).\\
\texttt{--min-contributors} & Minimum contributors (default: 1).\\
\texttt{--max-repos} & Cap on repositories processed.\\
\texttt{--max-functions} & Cap on extracted functions, applied after deduplication and stratified by repository.\\
\texttt{--max-classes} & Same cap, for classes.\\
\texttt{--engineered-only} & Apply the engineered-project filter.\\
\texttt{--[no-]deduplicate} & Enable or disable near-duplicate removal (default: on).\\
\texttt{--dedup-threshold} & Jaccard similarity threshold (default: 0.8).\\
\texttt{--export-format} & \texttt{jsonl} (default), \texttt{parquet}, or \texttt{both}.\\
\texttt{--output-dir} & Output directory (default: \texttt{./sieve\_output}).\\
\texttt{--github-token} & GitHub token (or \texttt{GITHUB\_TOKEN} variable).\\
\hline
\end{tabular}
\end{small}

\medskip
\noindent
Options without a flag, such as LLM-score annotation (\texttt{annotate\_llm\_score}), can be set only in the JSON file. The following command covers the scenario except for the scores: it discovers and filters the repositories, extracts the capped corpus without deduplication, and exports both formats:

\begin{lstlisting}[style=bashstyle]
sieve run \
  --language Python \
  --start-date 2024-06-01 --end-date 2025-06-01 \
  --min-stars 50 --min-contributors 5 \
  --max-repos 20 --max-functions 30 --max-classes 30 \
  --engineered-only --no-deduplicate --export-format both
\end{lstlisting}

\noindent
To also score the records, the same settings are written to a JSON file with \texttt{annotate\_llm\_score} set to true and run as \texttt{sieve run --config config.json}.

\textbf{Python API.} After installation, researchers import \texttt{SIEVEConfig} and \texttt{run\_pipeline}. The CLI parameters, together with the options that have no flag, are fields of a \texttt{SIEVEConfig} object, which is passed directly to \texttt{run\_pipeline}, enabling scripted or notebook-driven workflows. Listing~\ref{lst:sieve-api} configures and executes the same scenario. Deduplication is off by default in the API and is therefore not set, and the pipeline returns a summary dictionary with run statistics and the paths of the exported files.

\begin{lstlisting}[style=pythonstyle, numbers=left, numberstyle=\tiny\color{gray}, numbersep=5pt, xleftmargin=1.5em, caption={Example of using the SIEVE Python API.}, label={lst:sieve-api}]
from sieve.config import SIEVEConfig
from sieve.pipeline import run_pipeline

config = SIEVEConfig(
    language="Python",
    start_date="2024-06-01",
    end_date="2025-06-01",
    min_stars=50,
    min_contributors=5,
    max_repos=20,
    max_functions=30,
    max_classes=30,
    engineered_only=True,
    annotate_llm_score=True,
    export_format="both",
)

summary = run_pipeline(config)
print(summary["total_functions"], summary["output_paths"])
\end{lstlisting}

\begin{figure*}[t]
\centerline{\includegraphics[width=\textwidth]{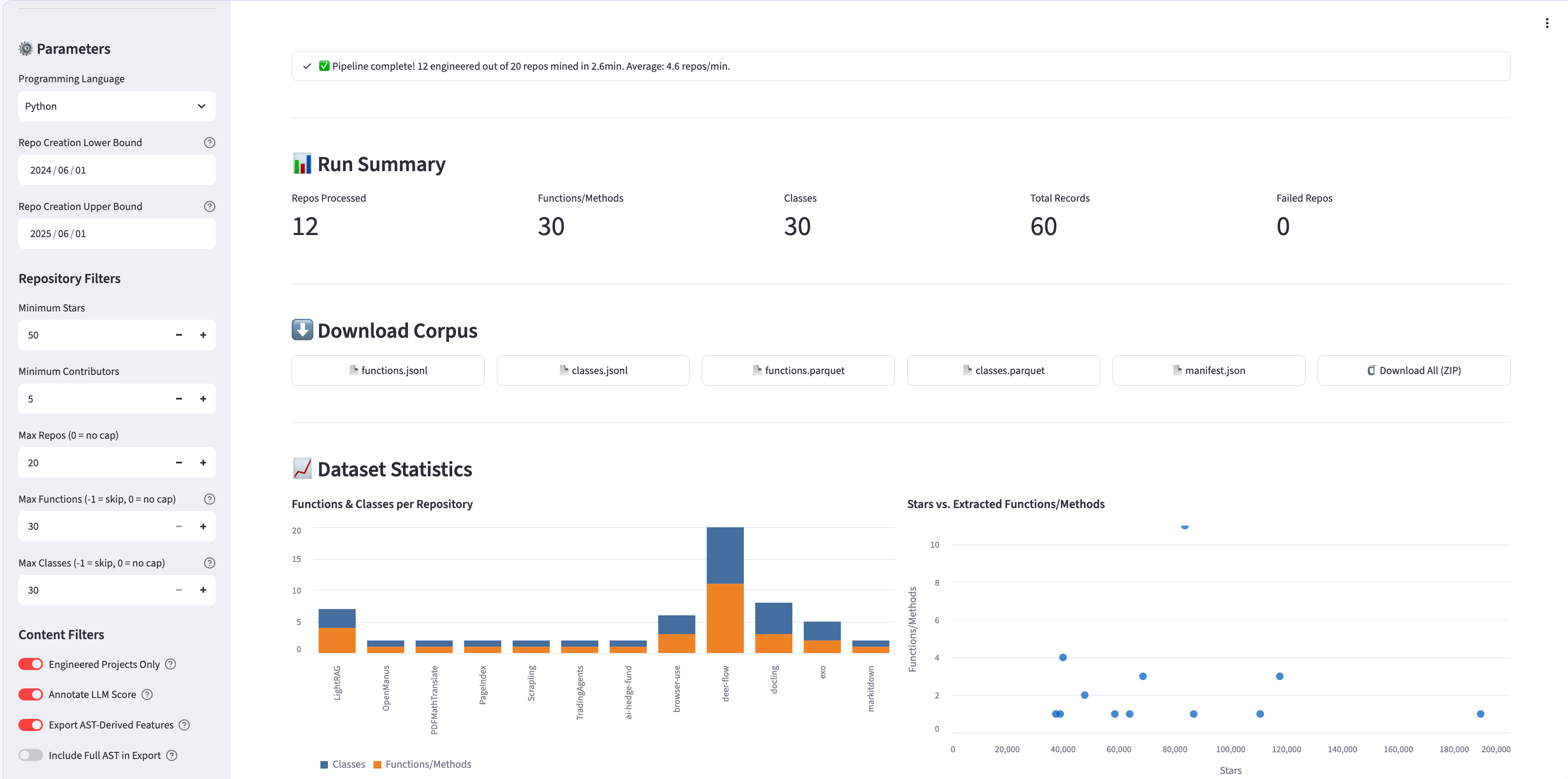}}
\caption{The \textsc{SIEVE} web interface after a pipeline run: 12 engineered
repositories processed from 20 discovered, yielding 60 records (30 functions,
30 classes) in 2.6 minutes.}
\label{fig:ui}
\end{figure*}

\textbf{Web interface.} The same fields are exposed as form controls in the web application, requiring no code. Fig.~\ref{fig:ui} shows the interface after a completed run with these settings, displaying the parameter configuration, pipeline log, run summary statistics, and corpus download options.
The corpus is downloaded in the format selected for the run (JSONL, Parquet, or both), as one file per granularity together with \texttt{manifest.json}, or as a single ZIP bundle.

\textbf{Deployment.} \textsc{SIEVE} is deployed as a Docker container on HuggingFace Spaces at \url{https://mrahman2025-sieve.hf.space}, so the web interface can be used without local installation. The CodeBERT classifier weights are hosted on HuggingFace Hub at \texttt{mrahman2025/sieve-llm-classifier}. A GitHub Actions pipeline runs 455 unit, integration, and end-to-end tests (83\% coverage) on every commit, with deployment to HuggingFace Spaces triggered only after the tests pass on the main branch.

% ── V. Evaluation ────────────────────────────────────────────────────────────
\section{Evaluation}
\label{sec:evaluation}
We evaluate \textsc{SIEVE} along three dimensions: the quality of the extracted corpus (Section~\ref{sec:validation}), the performance of the bundled LLM-generated-code classifier (Section~\ref{sec:classifier}), and a user survey (Section~\ref{sec:survey}).

\subsection{Corpus Quality Validation}
\label{sec:validation}

We conducted a manual quality validation of a stratified sample of 400 records drawn from the \textsc{SIEVE} extraction pipeline. The sample comprises 100 records per language (Python, Java, C++, JavaScript), equally split between function/method and class granularities (50 each). Each record was assessed independently across four dimensions: \textit{\textbf{Source Code Completeness (D1)}}: whether the source code is \textit{syntactically complete};~\textit{\textbf{Metadata Accuracy (D2)}}: whether metadata fields (e.g., function name, parameter list, method flag, method names) \textit{correctly reflect the source};~\textit{\textbf{Import Detection (D3)}}: whether the \textit{recorded imports} contain no false positives and no obvious false negatives; and~\textit{\textbf{Code Metrics (D4)}}: whether stored metrics \textit{match values recomputed independently} on the same source code. Each dimension was rated \textit{Correct} or \textit{Incorrect}, where partial issues count as incorrect. Accuracy is defined as: 

\begin{equation}
\text{Accuracy} = \frac{|\text{Correct}|}{N} \times 100
\label{eq:accuracy}
\end{equation}
where $|\text{Correct}|$ is the number of records rated \textit{Correct} on a dimension and $N$ is the number of records assessed in the group being scored (50 per language-granularity cell, 400 overall). Table~\ref{tab:validation} reports accuracy per dimension, language, and granularity, with 95\% confidence intervals from a cluster bootstrap~\cite{field2007bootstrapping} over repositories, which accounts for non-independence among records drawn from the same repository.

\begin{table}[t]
\caption{Validation accuracy (\%) per dimension, language, and granularity
(N=50 per cell). Brackets: 95\% cluster-bootstrap confidence intervals, shown below 100\%.}
\label{tab:validation}
\begin{center}
\begin{tabular}{llcccc}
\toprule
\textbf{Language} & \textbf{Granularity} & \textbf{D1\%} & \textbf{D2 \%} & \textbf{D3\%} & \textbf{D4\%} \\
\midrule
Python      & Function  & 100 & 100 & 100  & 100 \\
            & Class  & 100 & 100 & 100  & 100 \\
Java        & Function  & 100 & 96 [91--100] & 100 & 100 \\
            & Class  & 100 & 100 & 100 & 100 \\
C++         & Function  & 100 & 100 & 100 & 100 \\
            & Class  & 92 [91--100] & 100 & 100 & 100 \\
JavaScript  & Function  & 100 & 100 & 100  & 100 \\
            & Class  & 100 & 100 & 100  & 100 \\
\midrule
\textbf{Overall} & & \textbf{99.0} & \textbf{99.5} & \textbf{100.0} & \textbf{100.0} \\
\bottomrule
\end{tabular}
\end{center}
\end{table}

Table~\ref{tab:validation} shows that \textsc{SIEVE} extracts correct records in nearly all cases. Code metrics (D4) and import detection (D3) are correct for all 400 records, and 30 of the 32 language-granularity-dimension cells reach 100\%, including every Python and JavaScript cell. The only deviations are source completeness (D1) for C++ classes (92\%) and metadata accuracy (D2) for Java functions (96\%), which together account for 6 of the 1,600 ratings (400 records across four dimensions).

From Table~\ref{tab:validation}, D1 drops to 92\% for C++ classes. This drop is due to a known limitation of \texttt{tree-sitter-cpp}~\cite{githubGitHubNsumnercpptreesitter}: as its maintainers explain, C++ contains syntactic ambiguities that cannot be resolved without type information, so the grammar selects one parse by design~\cite{githubGrammarAmbiguous}. As a result, some extracted class snippets are parsed with errors even though the code itself is valid.

In contrast, the D2 drop to 96\% for Java functions originates in \textsc{SIEVE}'s extractor rather than in the grammar. Tree-sitter gives a method declared in a Java interface or enum the same node type as a method declared in a class, but our traversal recognizes only class declarations as enclosing scopes. Methods of other declarations are therefore recorded as ordinary functions instead of being attached to their parent. The same limitation applies to methods of C++ \texttt{struct}s, although it did not affect our sample.
Because scope handling is localized in each language's traversal function, these cases could be addressed in future versions, and the open-source release allows the community to contribute such extensions.

\subsection{LLM Classifier Performance}
\label{sec:classifier}

The objective of this evaluation is to measure how well the bundled classifier separates human-written from LLM-generated code, since its score (\texttt{llm\_score}) is attached to every extracted record. The approach follows prior work: GPTSniffer~\cite{nguyen2023snippet} fine-tunes CodeBERT~\cite{feng2020codebert} for this task, and our earlier study~\cite{rahman2024automatic} examines detection at function and class granularities. The classifier is a fine-tuned CodeBERT with a linear classification head, trained on 114,186 snippets (54,668 human-written and 59,518 LLM-generated) across Python, Java, JavaScript, and C++. Human code was mined with \textsc{SIEVE} from repositories created after January 2024, and LLM-generated code was produced by prompting nine LLMs (four for Python, five for the other languages) with signatures extracted from the human code. We evaluate on a held-out test set of 11,419 samples (a 10\% stratified split). The classifier achieves F1\,=\,0.9478, AUC-ROC\,=\,0.9902, precision\,=\,0.9257, and recall\,=\,0.9709, with per-language F1 between 0.9407 (C++) and 0.9515 (Java). Its higher recall than precision suits its role of flagging candidates for downstream filtering. The test set shares generators and prompting protocol with the training data, so these results do not establish generalization to unseen LLMs. Full training details and per-language results are documented on the \textit{About the Classifier} page of the \textsc{SIEVE} web application~\cite{sieveClassifierDocs}.

\subsection{User Survey}
\label{sec:survey}

We ran an anonymous online survey from September 1 to 30, 2026, and received 26 responses from SE researchers recruited through snowball sampling, so the response rate is unknown. Most respondents found configuring \textsc{SIEVE} easy or very easy (21 of 26), all 26 agreed that the extracted samples looked syntactically correct and complete, and 24 agreed that it saved them time. Twenty-five would use it in their own research (17 definitely, 8 with improvements), and all 26 agreed that contamination awareness matters for their work. The most requested improvements were more filtering options (8), faster execution (5), and more programming languages (4). These results are self-reported and come from a small, self-selected, anonymous sample, most of whose backgrounds we could not verify, so they indicate perceived usability and usefulness and not measured effort savings.

% ── VI. Tool Applicability ───────────────────────────────────────────────────

\section{Tool Applicability}
\label{sec:applicability}

\textbf{Researchers.} MSR studies on code typically chain separate tools for repository sampling, cloning, parsing, metric computation, dependency parsing, and classification, each with its own configuration and output format. \textsc{SIEVE} replaces this chain with one parameterized pipeline, a uniform record schema, and a manifest that records the configuration of every run, which lowers the effort of building and maintaining the tooling and the number of tools a researcher must learn. The same pipeline then supports different research threads: code quality and maintainability studies using the complexity, Halstead, and Maintainability Index metrics; dependency and library-usage studies using the extracted manifests and per-record imports; documentation studies using docstrings alongside source code; studies of inheritance and design using class records with parent classes, decorators, and method lists; comparisons of repository cohorts created in different periods, including the classifier scores of their code; and temporally bounded corpora for LLM evaluation. 

\textbf{Practitioners.} Engineering teams can use \textsc{SIEVE} for three concrete activities. For code quality assessment, the complexity, nesting, and Maintainability Index metrics computed over a corpus of comparable open-source projects give reference distributions against which a team can calibrate its own thresholds. For dependency analysis, the manifest lists the declared direct dependencies of each repository with their versions and kinds (main, development, or optional), showing which libraries and versions comparable projects adopt. For screening potentially AI-generated code, the bundled classifier can be applied to snippets through the Python API; its score is a probabilistic signal for prioritizing review, not evidence of authorship.

\textbf{Educators.} \textsc{SIEVE} can supply real-world material for three educational activities. For programming exercises, instructors can select functions and classes of a chosen language by size, complexity, or nesting depth. For code comprehension, records carry docstrings where present and classes carry skeletons of their signatures. For maintainability assessment, Maintainability Index, comment ratio, and complexity values let students inspect and refactor real code. These metrics describe structure and are only proxies for the difficulty a student experiences, so instructors should pair them with their own judgment. Restricting the creation-date range may reduce the chance that selected code already circulates as public solutions or appears in LLM training data, but it does not guarantee that the code is unseen by LLMs. 

% ── VII. Conclusion and Future Works ─────────────────────────────────────────
\section{Conclusion and Future Works}
\label{sec:conclusion}

We presented \textsc{SIEVE}, a parameterized GitHub corpus builder that gives SE researchers direct control over language, granularity, repository quality, and repository creation date when sourcing real-world code corpora. Rather than a static, pre-built dataset or an ad-hoc one-off script, it is reusable infrastructure for building the corpus a given study needs, supporting structural and maintainability analyses, dependency studies, class-level benchmarking, studies of AI-generated code, and LLM evaluation on temporally bounded corpora. A manual validation of 400 records across four languages and two granularities found accuracy between 99.0\% and 100\% on all four dimensions, with the two deviations traced to a C++ grammar limitation and to our extractor's handling of enclosing declarations, and the bundled classifier reaches an F1-score of 0.9478 on its held-out test set. The tool is open source under the MIT license and is available as a web interface, a command-line tool, and a Python API.

In future work, we want to extend language support beyond the current four, handle further enclosing declarations (interfaces, enums, and structs) in the extractor, replace the CodeBERT-based LLM-generated-code classifier with an interpretable alternative that can explain why a snippet is flagged rather than returning an opaque score, and address the improvements requested in our user survey, chiefly richer filtering options, faster execution, and additional languages. We hope \textsc{SIEVE} lowers the barrier for researchers to source well-annotated, real-world code corpora tailored to their study's needs, whatever those needs may be.

% ── Tool and Data Availability ────────────────────────────────────────────────────────────
\section*{Tool and Data Availability}
\label{sec:data-availability}
All artifacts are publicly available.
\begin{itemize}
    \item A live interactive demo can be found here~\cite{hfStreamlit}
    \item The GitHub repository can be found here~\cite{githubGitHubMrsumitbdsieve}
    \item LLM-generated code detection classifier weights can be found here~\cite{huggingfaceMrahman2025sievellmclassifierHugging}
\end{itemize}

% ── References ────────────────────────────────────────────────────────────────
\bibliographystyle{IEEEtran}
\bibliography{references}

\end{document}